\documentclass[a4paper,fleqn]{cas-dc}

\usepackage[authoryear]{natbib}   
\usepackage{amssymb}
\usepackage[utf8]{inputenc}
\usepackage{amsmath}
\usepackage{xcolor}
\usepackage[normalem]{ulem}
\usepackage{graphicx}
\usepackage{bm}
\usepackage{stfloats}
\usepackage{setspace}
\usepackage[switch]{lineno} 
\usepackage{soul}
\usepackage{caption}
\def\tsc#1{\csdef{#1}{\textsc{\lowercase{#1}}\xspace}}
\tsc{VP}
\tsc{NT}

\newcommand{\dt}[1]{\frac{\mathrm{d}#1}{\mathrm{d}t}}
\newcommand{\omg}{\bm{\omega}}
\newcommand{\omap}{\omega||\mathrm{MIA}}
\newcommand{\tomap}{o\omega||\mathrm{mMIA}}

\newcommand{\tax}{\bm{o}}
\newcommand{\taxa}{o}
\newcommand{\dv}{\,\mathrm{d}v}
\newcommand{\ivo}{\int_{v(t)}}

\begin{document}
\let\WriteBookmarks\relax
\def\floatpagepagefraction{1}
\def\textpagefraction{.001}
\shorttitle{Dynamic reorientation of tidally locked bodies}
\shortauthors{Pato\v{c}ka et al.}

\title [mode = title]{Dynamic reorientation of tidally locked bodies: application to Pluto}

\author[1]{Vojt\v{e}ch Pato\v{c}ka}[orcid=0000-0002-3413-6120]
\credit{Conceptualization, Methodology, Investigation, Visualization, Writing - Original draft preparation}
\author[1,2]{ Martin Kihoulou}[]
\credit{Conceptualization, Writing - Review \& Editing}
\author[1]{ Ond\v{r}ej \v{C}adek}[]
\credit{Conceptualization, Writing - Review \& Editing}

\address[1]{Charles University, Faculty of Mathematics and Physics, Department of Geophysics, V Hole\v{s}ovi\v{c}k\'{a}ch 2, 180 00 Prague, Czech Republic}
\address[2]{Laboratoire de Planétologie et Géosciences, Université de Nantes, 2 rue de la Houssinière, 44322 Nantes, France}

\begin{abstract}
Planets and moons reorient in space due to mass redistribution associated with various types of internal and external processes. While the equilibrium orientation of a tidally locked body is well understood, much less explored are the dynamics of the reorientation process (or true polar wander, TPW, used here for the motion of either the rotation or the tidal pole). This is despite their importance for predicting the patterns of TPW-induced surface fractures, and for assessing whether enough time has passed for the equilibrium orientation to be reached. The only existing, and relatively complex numerical method for an accurate evaluation of the reorientation dynamics of a tidally locked body was described in a series of papers by \citet{Hu2017a,Hu2017b,Hu2019}. Here we demonstrate that an identical solution can be obtained with a simple approach, denoted as $\tomap$, because, contrary to previous claims, during TPW the tidal and the rotation axes closely follow respectively the minor and the major axes of the total, time-evolving inertia tensor. Motivated by the presumed reorientation of Pluto, the use of the $\tomap$ method is illustrated on several test examples. In particular, we analyze whether reorientation paths are curved or straight when the load sign and the mass of the host body are varied. When tidal forcing is relatively small, the paths of negative anomalies (e.g.~basins) towards the rotation pole are highly curved, while positive loads may reach the sub- or anti-host point in a straightforward manner. Our results suggest that the Sputnik Planitia basin cannot be a negative anomaly at present day, and that the remnant figure of Pluto must have formed prior to the reorientation. Finally, the presented method is complemented with an energy balance that can be used to test the numerical solution and to quantify the changes of orbital distance due to the reorientation. A new release of the custom written code LIOUSHELL that is used to perform the simulations is made freely available on GitHub.

\end{abstract}

\begin{highlights}
\item Dynamic reorientation of a tidally locked body is obtained with a simple method. 
\item A negative load may easily change its longitude when the tidal bulge is small.
\item It is unlikely that Sputnik Planitia formed before Pluto's elastic lithosphere.
\end{highlights}


\maketitle

\section{Introduction}

Planets and moons reorient with respect to the stars when internal processes or external impacts change their distribution of mass. The phenomenon was first analyzed for Earth, and is thus referred to as the ``true'' polar wander (TPW), to be distinguished from the ``apparent'' motion of the rotation pole that is perceived by an observer on a drifting continent \citep[e.g.][]{Besse2002}. Measuring the Earth's rotation pole and studying its dynamics has a long history \citep{Munk1960}, but for planets and moons with a non-negligible tidal bulge the approach to TPW is usually quite crude: only the equilibrium orientation is assessed. That is, the equilibrium inertia tensor of the body is evaluated and diagonalized, with the principal directions marking the final positions of the rotation and tidal axes \citep[for a review, see][]{Matsuyama2014}. 

While the dynamics of reorientation of a tidally locked body are governed by the viscoelastic readjustment of the rotational and tidal bulges, in the equilibrium state the symmetry axes of the bulges are by definition aligned with respectively the rotational and tidal axes. Therefore, the only constituents of the inertia tensor that need to be determined in order to get the principal directions in the equilibrium state are the inertia tensor of the load, and that of the so-called remnant figure. The remnant (or fossil) figure forms when the hydrostatic shape of a planet, that is fluid across its entire depth, ``freezes'' into the growing lithosphere as the planet cools. If the centrifugal or tidal forces later change in response to TPW, the presence of an elastic lithosphere prevents the body from reaching the hydrostatic equilibrium again. Therefore, the fossil figure always reflects the primordial rotation and tides stays always aligned with the primordial directions of rotation and tides, and counterbalances the load that drives TPW \citep{Willemann1984}.

The equilibrium approach neglects the dynamics of reorientation -- it is concerned only with the final orientation of the body. Nevertheless, it can be used to estimate the TPW path when the investigated load is treated as a sequence of loads with a gradually increasing amplitude. The equilibrium orientation is then computed for each of these partial loads, yielding the evolution of reorientation, known as the fluid-limit solution \citep[e.g.][]{Keane2016}. The assumption behind this approach is that the load formation is slow when compared to the rotational and tidal bulge readjustment.

Several planets and moons are thought to have reoriented in the past, with their most striking surface features located near the poles or the equator, depending on whether the associated gravity anomaly has a negative or a positive sign \citep[e.g.][]{Keane2014,Bouley2016,Nimmo2016,Tajeddine2017}. However, the dynamic feasibility of these hypothesised, often large-angle reorientations is only rarely assessed or put in the context of the thermal and orbital history of the body. Moreover, the loading itself may depend on insolation and thus on the orientation of the body in space \citep{Ojakangas1989}. In such a case, it becomes crucial to compute the viscoelastic response to the loading as well as the TPW dynamics in a self-consistent way using a single numerical model.

Reorientation scenarios are often supported by an analysis of the surface stress patterns \citep[e.g.][]{Tajeddine2017,Keane2016}. When the direction of the centrifugal or tidal forces changes, the lithosphere is subjected to stress that is manifested by changes in the tectonic pattern. A large reorientation can thus generate a global network of extensional and compressional fractures, depending on the position of each point with respect to the old and the new centrifugal and tidal potentials \citep{Melosh1980}. In order to predict the map of tectonic stresses, the entire TPW is typically treated as a single, instantaneous event \citep[e.g.][]{Nimmo2016}, or as a progression of fracture-forming events \citep[e.g.][]{Keane2016}. Using the fluid-limit framework is somewhat paradoxical in this regard: the TPW is assumed to be slow so that the time needed for the bulge readjustment is negligible, but at the same time, reorientation is considered to be instantaneous for the purpose of surface stress evaluation (or step-wise in the case of the progression of events).

In a series of papers, \citet{Hu2017a,Hu2017b,Hu2019} developed the first method for computing the dynamics of reorientation of a tidally locked body, and tested it against the fluid-limit approach. The method was also compared with the so-called quasi-fluid approximation, which is traditionally used to evaluate the TPW on Earth \citep{Lefftz1991,Ricard1993} and in which the rotational bulge readjustment is simplified by neglecting all short-term relaxation modes. The TPW speed computed by \citet{Hu2019} lies in between the fluid-limit and the quasi-fluid solutions, and all the three methods converge when TPW is much slower than the bulge readjustment. 

The algorithms developed by \citet{Hu2017a,Hu2017b,Hu2019} are relatively complex. At each time step, the problem is transformed to the ``bulge-fixed'' frame, in which the linearized Liouville equations (LLE) are solved. For a tidally locked body, the LLE are solved separately for the tidal and for the rotation vectors, and the condition of perpendicularity of the vectors is achieved by an iterative adjustment of the obtained LLE solutions \citep[see Section 2.2 and Appendix A in][]{Hu2019}.

In cases without tidal forcing, \citet{Patocka2021} obtained the same results as \citet{Hu2017a,Hu2017b} with a simple method, denoted as $\omap$. The rotation axis coincides with the major (or main) inertia axis throughout the entire TPW simulation, which has been overlooked by \citet{Hu2017a,Hu2017b}, perhaps because the $\omap$ assumption was originally linked to the quasi-fluid approximation. However, the two simplifications in the governing equations, namely the $\omap$ approximation of the conservation of angular momentum and the quasi-fluid approximation of the viscoelastic response to loading, should be treated separately \citep{Patocka2021}. While the short-term relaxation modes must be handled with care \citep{Hu2017a,Hu2017b,Hu2019}, the $\omap$ assumption is generally valid \citep{Patocka2021}.

Here, the $\omap$ method is extended to encompass tidally locked bodies. We show that the solutions from \citet{Hu2019} can be reproduced when it is assumed that the tidal and rotation axes coincide with the minor (mIA) and major (MIA) axes of the total inertia tensor, respectively. This allows for a simple and robust method for computing the reorientation of synchronously rotating planets and moons, hereafter denoted as $\tomap$.


One general question to ask is whether a tidally locked body is likely to reach its equilibrium orientation along the shortest possible path, and how this path depends on the size of the host body. \citet{Hu2019} argue that the reorientation path tends to straighten as the relative size of the tidal bulge reduces, being nearly straight when the tidal/rotational bulge ratio drops to about $0.1$. The $\tomap$ assumption, on the other hand, indicates that even a tiny tidal bulge could substantially distort the TPW paths of tidally locked bodies. This is because it is easy for the load to move the minor and the intermediate inertia axes along the equatorial plane when the tidal bulge is small, making the reorientation path highly curved in effect. 

This problem becomes particularly interesting when the reorientation of Pluto due to the formation of the Sputnik Planitia basin is addressed. Owing to the relatively small mass of Charon, the tidal bulge of Pluto is much smaller than its rotational bulge. Centered at 176°E 24°N, the Sputnik Planitia basin is thought to be a positive gravity anomaly despite its negative topography \citep{Keane2016}, suggesting a subsurface ocean \citep{Nimmo2016}. 

Preceded by a description and validation of our method in sections \ref{sec:method} and \ref{sec:triton}, the reorientation of Pluto is investigated in sections \ref{sec:pluto} and \ref{sec:discussion}. In section \ref{sec:energy}, the $\tomap$ method is analyzed from the point of view of the energy conservation law. The main findings of the study are summarized in section 7.

\section{The $\tomap$ Approximation}\label{sec:method}

The deformation of a hydrostatically prestressed incompressible viscoelastic ice shell is calculated by integrating the conservation equations for mass and momentum and the constitutive law for a Maxwell body \citep{Tobie2008,Patocka2018,Patocka2021}. The outer boundary is treated as a quasi-free surface \citep[Eq.~7 in][]{Patocka2021}, while the bottom boundary is assumed to be in contact with an inviscid fluid that is in hydrostatic equilibrium and rotates synchronously with the shell. In section \ref{sec:pluto}, where Pluto is investigated, the interior water ocean surrounds a silicate core. In this case, we assume that the ice shell and the ocean form a coupled system that reorients simultaneously, while the core remains fixed with respect to the rotational and tidal axes and its contribution to the inertia tensor can thus be omitted \citep[cf.][]{Ojakangas1989}.

The numerical method that we use to compute the deformation is detailed and validated in \citet{Patocka2018} and \citet{Patocka2021}. The only difference here is that the centrifugal potential $\Psi$ is replaced by $\Psi + \Theta$, where $\Theta$ represents the tidal potential. While the centrifugal potential can be written as
\begin{linenomath*}
\begin{equation}\label{eqRotpot2}
    \Psi = \frac{1}{2}\left( (\omg\cdot\bm{r})^2 - \omega^2 r^2 \right),
\end{equation}
\end{linenomath*}
where $\omg$ is the angular velocity and $\bm{r}$ is the position vector, the tidal potential takes the following form:
\begin{linenomath*}
\begin{equation}\label{eqTidpot}
    \Theta = \frac{1}{6}\taxa^2 r^2 - \frac{1}{2}(\tax\cdot\bm{r})^2 .
\end{equation}\end{linenomath*}
For convenience, we have introduced a ``tidal vector'' $\tax$ that points toward the host body and has the magnitude:
\begin{linenomath*}\begin{equation}\label{eqTidamp}
\taxa = \frac{3\,GM_\mathrm{h}}{a^3},
\end{equation}\end{linenomath*}
where $G$ is the universal gravitational constant and $a$ is the distance from the host body of mass $M_\mathrm{h}$. 

For a tidally locked planet or moon of mass $M$, the angular frequency $\omega$ is equal to the orbital frequency,
\begin{linenomath*}\begin{equation}\label{eqTidlock}
\omega = \sqrt{\frac{G(M_\mathrm{h}+M)}{a^3}} = \taxa \sqrt{\frac{M_\mathrm{h}+M}{3M_\mathrm{h}}}.
\end{equation}\end{linenomath*}
Comparison of Eqs \eqref{eqRotpot2} and \eqref{eqTidpot} shows that the centrifugal and tidal potentials have the same degree two structure, and differ only in the sign, amplitude, and the degree zero component. The torque of the centrifugal force can be expressed as,
\begin{linenomath*}\begin{multline}
\ivo{\bm{r}\times(-\nabla \Psi)\,\rho\dv} = \ivo{\bm{r}\times(\bm{r}\,\omega^2 - \omg\,(\omg\cdot\bm{r}))\,\rho\dv} \\ 
= \omg\times\ivo{\bm{r}\,(\omg\cdot\bm{r}) \,\rho\dv} = -\omg\times (\bm{I}\cdot\omg). \label{eqRotmoment}
\end{multline}\end{linenomath*}
Here $v(t)$ is the volume of the body at time $t$, $\rho$ is the density, and $\bm{I}$ is the inertia tensor,
\begin{linenomath*}\begin{equation}
    \bm{I} = \ivo{((\bm{r}\cdot\bm{r})\, \mathbb{1} - \bm{r}\otimes\bm{r})\,\rho\dv}, \label{eqIdef}
\end{equation}\end{linenomath*}
where $\mathbb{1}$ is the identity tensor. The eigenvalues of $\bm{I}(t)$ are referred to as $C, B$, and $A$, in order of their decreasing value. In Eq.~\eqref{eqRotmoment} we employed the fact that a part of $\bm{I}\cdot\omg$ is parallel to $\omg$ and thus does not contribute to the cross product. Using the same procedure as above, we can express the torque of the tidal force, denoted as $\bm{M}$, 
\begin{linenomath*}\begin{equation}\label{eqTidmoment}
   \bm{M} = \ivo{\bm{r}\times(-\nabla \Theta)\,\rho\,\dv} = \tax\times (\bm{I}\cdot\tax).
\end{equation}\end{linenomath*}
Therefore, the Liouville equation (LE) can be expressed as follows:
\begin{linenomath*}\begin{equation}\label{eqFullLE}
    \dt{} (\bm{I}\cdot\omg) + \omg\times(\bm{I}\cdot\omg) = \tax\times (\bm{I}\cdot\tax) .
\end{equation}\end{linenomath*}
Note that we formulate the LE in the Tisserand (or body-fixed) frame \citep{Munk1960}.

In previous studies, the time derivative in the LE was linked to the short-term viscoelastic relaxation modes of planetary mantles. The problem of TPW was solved using the so-called quasi-fluid approximation, in which the time derivative in the LE and the relaxation modes that are fast compared to TPW were omitted \citep[i.e., the quasi-fluid approximation, see][]{Lefftz1991,Ricard1993}. In the absence of tidal forcing, the LE was thus used in the form $\omg\times(\bm{I}\cdot\omg) = \bm{0}$. The validity of this approximation was questioned by \citep{Hu2017a,Hu2017b}, who showed that omitting the short-term relaxation modes is a potential source of error in the TPW solutions. Based on this finding, the authors assumed that the time derivative in the LE cannot be neglected, and developed a sophisticated algorithm that solves the LLE in a computational frame whose $z-$axis advances along with $\omg$ \citep[see section 3.2 in][]{Hu2017a}. However, as shown by \citet[][Appendix A]{Patocka2021}, there is no strict connection between the time derivative in the LE and the short-term viscoelastic relaxation modes. When the shell relaxation is completely resolved, i.e.~when the employed maxwellian rheology includes all the relaxation modes, then the accuracy of the TPW solution is not harmed by employing the simple formula $\omg\times(\bm{I}\cdot\omg) = 0$ \citep{Patocka2021}. In other words, the source of error lies only in computing the time evolving shape of the shell inaccurately, not in dropping the time derivative term from the LE.

Motivated by the above, we write Eq.~\eqref{eqFullLE} as
\begin{linenomath*}\begin{equation}\label{eqtomap}
    \omg\times(\bm{I}\cdot\omg) = \tax\times (\bm{I}\cdot\tax).
\end{equation}\end{linenomath*}
A straightforward solution of Eq.~\eqref{eqtomap} is to place the rotation vector $\omg$ along the major axis of inertia, MIA, and the tidal vector $\tax$ along the minor axis of inertia, mIA. The orientation of the body as a function of time can thus be obtained by diagonalizing the inertia tensor $\bm{I}(t)$ at each time step. This special solution does not help in determining the magnitudes of $\omg$ and $\tax$, but note that it balances the LHS and RHS of Eq.~\eqref{eqtomap} by setting both sides to zero -- it therefore satisfies also the full LE, Eq.~\eqref{eqFullLE}, provided that $C\omega = \mathrm{const}$ (with $C$ being the time-evolving major moment of inertia and $\omega$ the spin rate). While Eq.~\eqref{eqtomap} governs the directions of the vectors $\omg$ and $\tax$ (the orientation of the body), the $C\omega = \mathrm{const}$ condition can be used to describe the changes in the spin rate. In summary, we define the $\tomap$ method as the following set of equations: 
\begin{linenomath*}\begin{align}
    \omg\times(\bm{I}\cdot\omg) = 0, \label{eqtomap_1}\\
    \tax\times (\bm{I}\cdot\tax) = 0, \label{eqtomap_2}\\
    \dt{} (\bm{I}\cdot\omg) = 0. \label{eqtomap_3}
\end{align}\end{linenomath*}

In sections \ref{sec:triton}, \ref{sec:pluto}, and \ref{sec:discussion} we perform numerical simulations of planetary reorientation. In these sections, only the directions of $\omg$ and $\tax$ are of interest. The therein reported results could be reproduced to a high degree of precision even if the changes in the spin rate were disregarded, that is, if one simply assumed $\omega = \mathrm{const}$ instead of solving Eq.~\eqref{eqtomap_3}. However, as explained in detail in section \ref{sec:energy}, Eq.~\eqref{eqtomap_3} is crucial in the energy balance of TPW, and must be included when an energetically consistent formulation is desired.

Note that Eqs \eqref{eqtomap_1} and \eqref{eqtomap_2} resemble the fluid-limit approach. The difference is that we treat $\bm{I}$ as a time dependent quantity that accounts for the existence and continuous readjustment of the rotational and tidal bulges, while in the fluid-limit approach $\bm{I}$ is just the final ($t{\rightarrow}\infty$) contribution of the load combined with that of the fossil figure. Our $\tomap$ method exploits the simplicity of the fluid-limit formulation, but it is, as shown below, equivalent to the fully dynamical method of \citet{Hu2017a}. Thus, we can capture complex, time-dependent processes, without making assumptions about the relative speed of the bulge readjustment and the TPW rate.

One way to understand the $\tomap$ method is that it filters the free oscillations of the body. This is because wobbling is the situation in which $\omg$ and MIA have different directions, making the $\omg\times(\bm{I}\cdot\omg)$ term non-zero, and thus triggering a non-zero rate of the angular momentum $\bm{I}\cdot\omg$ as measured in the body-fixed frame, the increment of $\bm{I}\cdot\omg$ being perpendicular to $\omg$ and periodically revolving around $\bm{I}\cdot\omg$ \citep[see e.g.~Figs~2 and 5 in][]{Spada1996b}. Avoiding the wobble, whose characteristic time scale is typically short when compared to geological time scales, is a big advantage in terms of computational efficiency, but has an obvious downside: our method cannot be applied to slowly-rotating bodies such as Venus, on which mega-wobble is thought to be the dominant reorientation mechanism \citep{Spada1996b}. On bodies other than Venus, free oscillations are typically assumed to be only a small perturbation that can be linearly combined with the secular motion of the pole \citep[e.g.][]{Martinec2014}, eventually decaying to zero \citep{Nakada2012} unless having a continuous excitation source \citep{Gross2000}.

\section{Validation: Test Case}\label{sec:triton}

\begin{figure*}[t]
    \centering
    \includegraphics[width=1.0\textwidth]{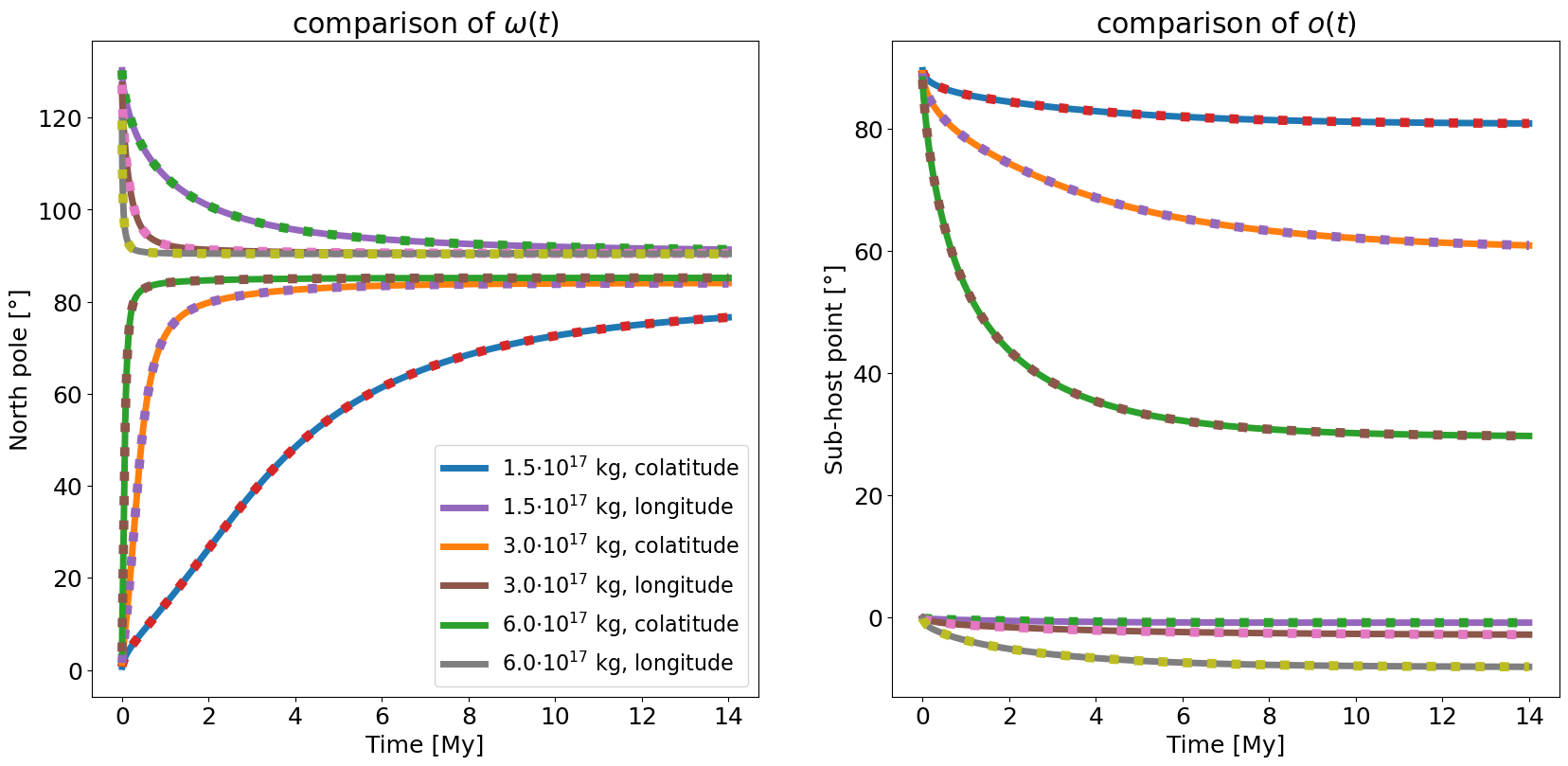}
    \includegraphics[width=1.0\textwidth]{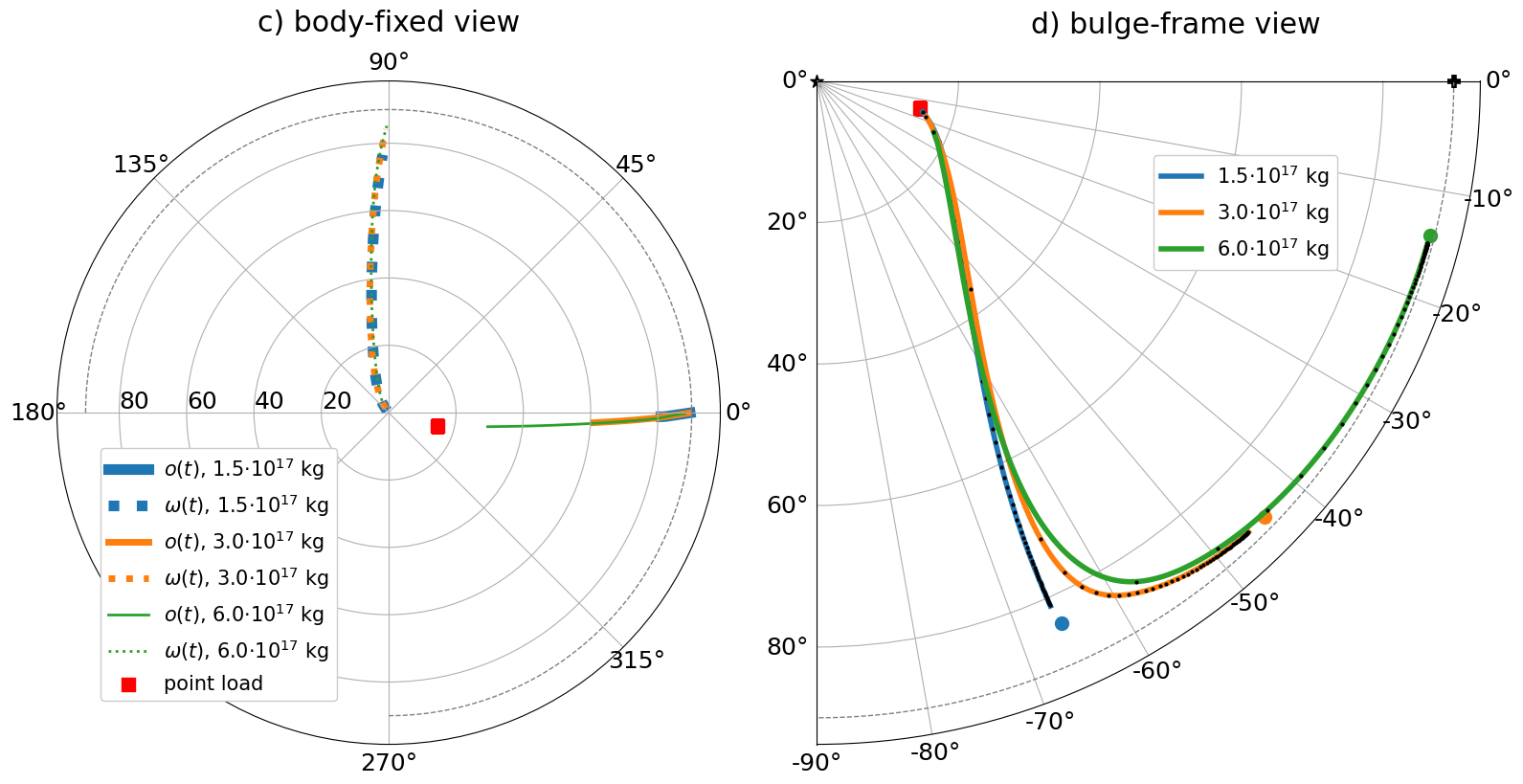}
    \caption{The test case of Triton. In panels a) and b), the colatitude and longitude (dotted lines) of respectively $\omg$ and $\tax$ as obtained in the body-fixed frame are plotted as a function of time and compared over the solutions from the top panel of Fig.~1 in \citet{Hu2019} (solid lines). In panel c), the simulation results are depicted in polar coordinates. In panel d), the same results are plotted in the bulge-fixed frame, in which the rotational axis intersects the surface of the body at colatitude 0°N at all times (black star), while the sub-host point is at all times located at longitude 0° and colatitude 90° (black cross). The red square indicates the initial position of the loads, while the coloured circles mark the equilibrium positions. Temporal evolution is illustrated by the small black dots that evenly sample the trajectory of each load in time intervals of 250 ky.}
    \label{fig:Triton}
\end{figure*}

First, we compare the $\tomap$ solution over the algorithm of \citet{Hu2019}. We perform a series of simulations for a model Triton, described in Table 1 and Fig.~1 of \citet{Hu2019}. The tidally locked moon is loaded with a point mass of respectively 1.5, 3.0, and 6.0 $\cdot 10^{17}$ kg, placed at 15° colatitude and -15° longitude. The ice shell includes a 10 km thick elastic lithosphere that formed after the body had reached the hydrostatic shape. The time evolution of colatitude and longitude of $\omg(t)$ and $\tax(t)$ computed with the $\tomap$ method (Fig.~\ref{fig:Triton}a,b, dotted lines) matches perfectly the original solution published by \citet{Hu2019} (solid lines). Despite its simplicity, Eq.~\eqref{eqtomap} is thus equivalent to the algorithms developed by \citet{Hu2019}.

In Fig.~\ref{fig:Triton}c, the solutions shown in panels a and b are plotted on the surface of the globe. This graphic representation allows the evolution of $\omg(t)$ and $\tax(t)$ to be shown with a single line each. On the other hand, it does not provide a clear view of how the reorientation progresses in time.

The solutions in panels a-c are plotted in the body-fixed (Tisserand) frame, i.e., in the frame in which the governing equations are formulated and solved. A more convenient way to visualize reorientations of tidally locked bodies was proposed by \citet{Hu2017a}, who used the so-called bulge-fixed frame. In this frame, the rotation axis always intersects the surface of the body at colatitude 0°N, while the sub-host point (tidal axis) is always located at longitude 0° and colatitude 90° (Fig.~\ref{fig:Triton}d). In this representation, the lines show how the colatitude and longitude of the load change in time (i.e., the lines show the time-evolving geographic position of the load).

\begin{figure}
    \centering
    \includegraphics[width=0.45\textwidth]{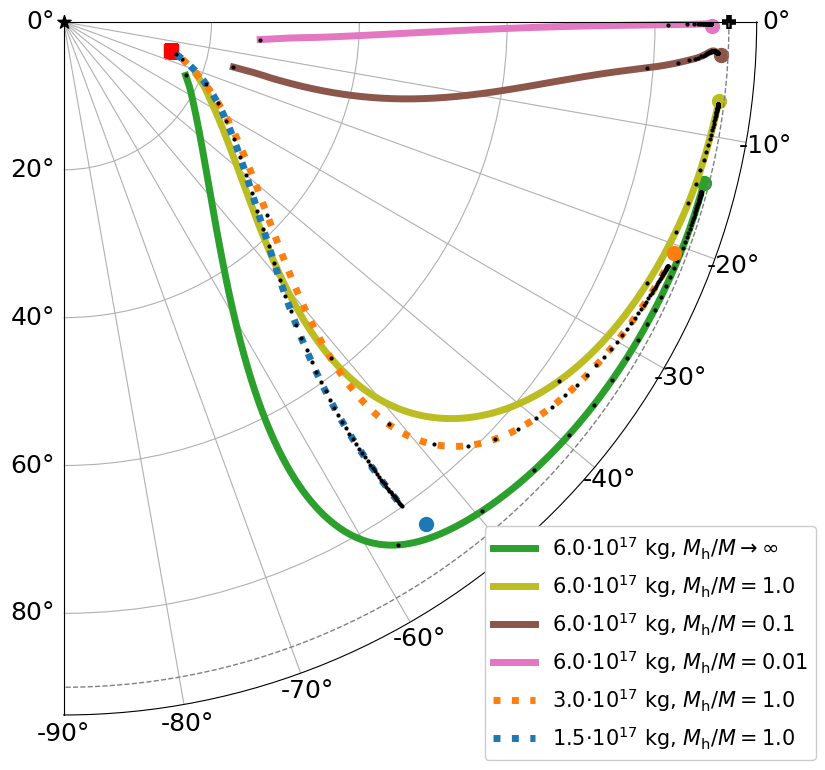}
    \caption{The influence of $M_\mathrm{h}/M$ on the path of a positive load, depicted in the bulge-fixed frame (cf.~Fig.~\ref{fig:Triton}d). Temporal evolution is illustrated by the small black dots that are evenly sampled in time at intervals of 250 ky. The load path straightens when $M_\mathrm{h}/M$ or the load size decrease.}
    \label{fig:MhMinfluence}
\end{figure}

In the three simulations presented in Fig.~\ref{fig:Triton}, Triton is assumed to orbit a planet much heavier than itself ($M_\mathrm{h}/M\rightarrow \infty$ in Eq.~\eqref{eqTidlock}). Following the exercise in Fig.~2 of \citet{Hu2019}, we perform additional simulations with a point load of mass $6\cdot 10^{17}$ kg, this time varying the $M_\mathrm{h} / M$ ratio, to investigate a system in which the mass of the host body is comparable with or even smaller than the mass of the studied body ($M_\mathrm{h} / M = 1, 0.1,$ and $0.01$, see Fig.~\ref{fig:MhMinfluence}).

If $M_\mathrm{h}/M\rightarrow \infty$, the position of the rotation axis changes by almost 90° in less than 2 Myr for the heaviest load, while the sub-host point takes longer to readjust (green lines in Fig.~\ref{fig:Triton}a and b). As a result, the load path is highly curved. This is not obvious in the body-fixed frame (Fig.~\ref{fig:Triton}c), but becomes apparent when the solution is plotted in the bulge-fixed frame (green line in Fig.~\ref{fig:Triton}d). As the $M_\mathrm{h} / M$ ratio is decreased, the time evolution of $\omg$ and $\tax$ becomes more balanced, and the equilibrium is reached in a more straightforward manner. Similarly, the load path straightens when $M_\mathrm{h} / M$ is constant and the load amplitude is decreased, but this is mainly because the final (equilibrium) position of the load changes significantly (dotted lines in Fig.~\ref{fig:MhMinfluence}).

In the next section, we apply our method to Pluto to simulate the ancient reorientation due to the Sputnik Planitia basin. The ``host'' body in this case is Charon, corresponding to $M_\mathrm{h} / M \approx 0.1$. For this mass ratio, the motion of a positive load occurs along a nearly straight line in the bulge-fixed frame (Fig.~\ref{fig:MhMinfluence}, brown curve). Since the sign of the load on Pluto is still a subject of debate, we also investigate the case where the load is negative and then compare the results of our simulations with observations.

\section{Reorientation of Pluto}\label{sec:pluto}

The topography of Pluto is dominated by a deep basin known as Sputnik Planitia. Its elliptic shape and multiring structure indicate an impact origin \citep{McKinnon2017}, but the basin could also have formed by the weight of accumulated nitrogen ice \citep{Hamilton2016}.

The geographic location of Sputnik Planitia (24°N, 176°E) suggests that it is a positive gravity anomaly, balanced ca.~25° away from the anti-Charon point by the remnant figure of Pluto \citep{Keane2016}. In order to explain how such a deep basin could increase the local gravity, \citet{Nimmo2016} speculate that there must be a dense interior ocean below the ice shell. One problem with this scenario is that impact simulations indicate that the post-impact ocean uplift is not sufficiently large to compensate the negative surface topography, and it would soon disappear due to the low viscosity of ice near the ice-ocean boundary \citep{Johnson2016}. Both \citet{Keane2016} and \citet{Nimmo2016} focus on the proximity of the basin center to the tidal axis, but they take little account of the fact that the center is very close to the 180° meridian, i.e., to the plane which contains both the tidal and rotational axes.

In the study of \citet{Hamilton2016}, on the other hand, Sputnik Planitia is shown to have reached 180° longitude already during the tidal despinning of Pluto caused by its large satellite Charon, and the latitude of the nitrogen deposit that presumably formed the basin (24°N) is shown to agree well with the band of low insolation of the highly tilted Pluto. However, the reorientation of the dwarf planet is addressed in a qualitative rather than a quantitative way, and it is not clearly explained how the basin could stay away from the equator without being balanced by a fossil rotational bulge. Since the authors argue that Sputnik Planitia was formed early in Pluto's history, it is unlikely that such a fossil bulge could have existed.

Finally, it was proposed by \citet{Kihoulou2022} that, regardless of its origin, Sputnik Planitia could have reoriented Pluto's primordial, thin ice shell such that its center was at the anti-Charon point. As Pluto cooled and the crustal thickness increased, the sign of the basin gravity anomaly changed, resulting in a further reorientation stage, during which the load began to drift northward. Assuming a 50 km thick elastic lithosphere (i.e., comparable to that assumed by \citet{Keane2016} and \citet{Nimmo2016}), \citet{Kihoulou2022} showed that this motion stops when the load reaches a latitude of about 25°, corresponding to the present position of Sputnik Planitia. It should be mentioned, however, that dynamics of the second reorientation stage were simplified in that the initial position of the load was fixed at the longitude of 180° and the stability of the solution with respect to small perturbations of the initial position was not tested.

The TPW is controlled by the viscoelastic relaxation of the ice shell occurring in response to changes in the rotational and tidal potentials. We assume that the shell is 150 km thick \citep{Johnson2016} and its viscosity varies with radius as follows \citep{Goldsby2001}:
\begin{linenomath*}\begin{equation}\label{eqicevisc}
    \eta_\mathrm{ice} = \frac{T(r)\, d_\mathrm{grain}^2}{ 3 A_\mathrm{act} } \exp\left(\frac{E_\mathrm{act}}{R_\mathrm{gas} T(r)}\right) ,
\end{equation}\end{linenomath*}
where $T(r)$ is the conductive temperature profile, $R$ is the universal gas constant, $d$ is the grain size and $A$ and $E$ are diffusion creep parameters (see Table 1). Temperature of the outer surface is 47 K and the ice/water interface is at 265 K (accounting for the pressure dependence of the melting temperature). A viscosity cutoff of $10^{24}$ Pa s is imposed for numerical reasons. Due to the low surface temperature of Pluto, most of the shell is governed by the cut-off viscosity. While the characteristic time scale of TPW strongly depends on the viscosity of ice and the ice shell thickness, it can be demonstrated that the load path is much less affected by the choice of these parameters. Since the relaxation time is not of primary importance for the discussion of Pluto's reorientation, we prescribe $\eta_\mathrm{max}$ rather arbitrarily, assess the plausibility of the above hypotheses by analyzing the load paths in the bulge-fixed frame, and discuss the robustness of our results only later in section \ref{sec:discussion}.    
   
We investigate the TPW induced by a disc load, whose thickness is increasing linearly from zero to the listed value over the time $t_\mathrm{load}$ (Table 1). As in the case of the point-load studied in section \ref{sec:triton}, the spherical disc is treated as a fixed contribution to the inertia tensor, i.e., no compensation of the load over time is considered. Note than when an impact basin is formed, it is a priori unclear whether the sign of the gravity anomaly it creates is positive or negative \citep[see, e.g., the Extended Data Figure 1 in][]{Keane2016}. In the absence of data, choosing a representative load amplitude is problematic. We investigate spherical discs of thicknesses ranging from -300 to 300 m, allowing us to capture the possible scenarios.

\begin{figure*}[t]
    \centering
    \includegraphics[width=0.45\textwidth]{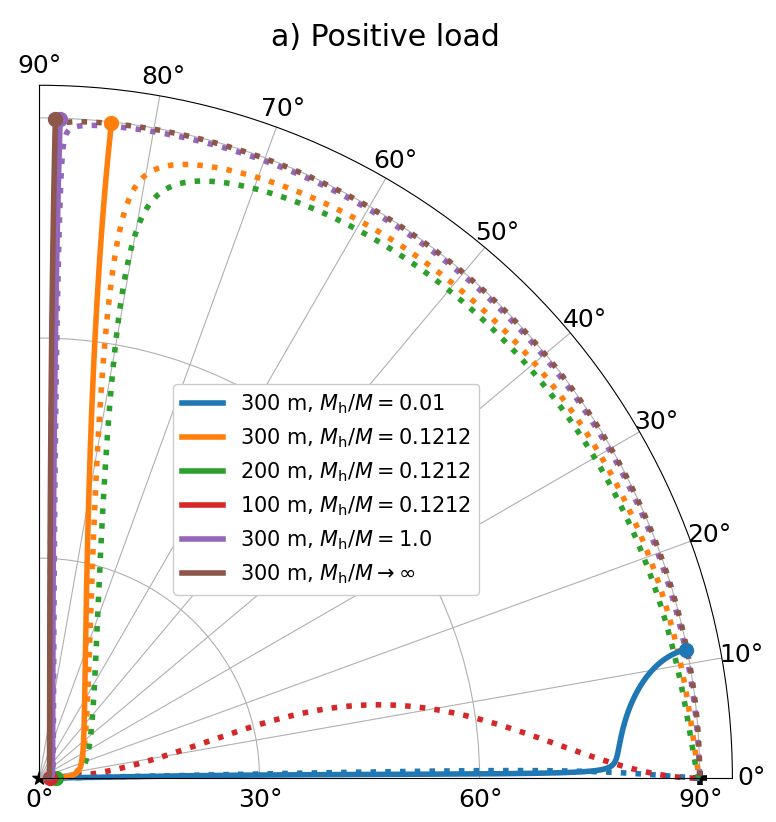}
    \includegraphics[width=0.45\textwidth]{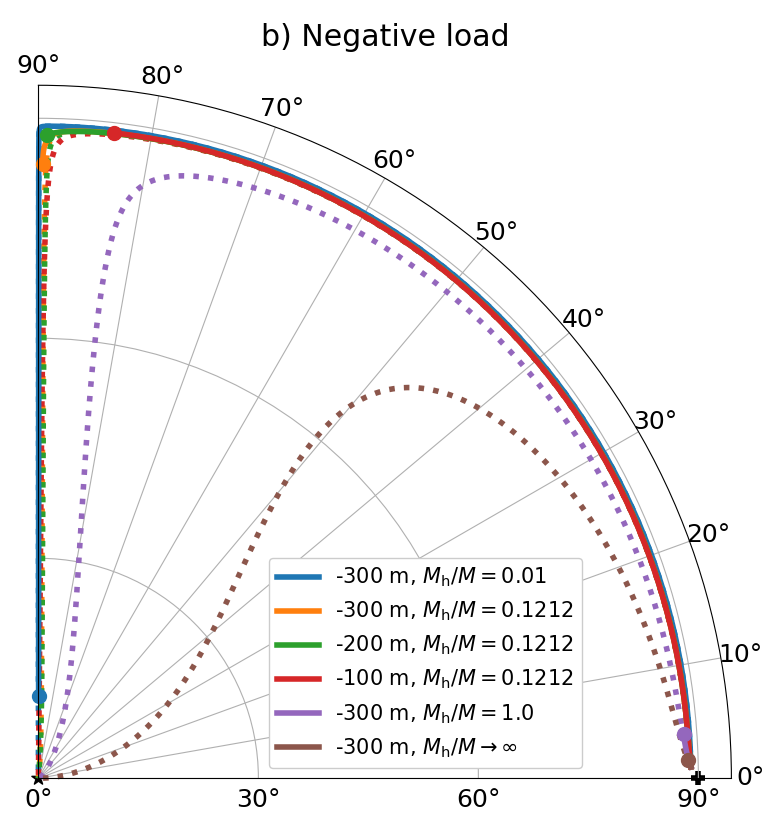}
    \caption{Load paths on a satellite (Table \ref{tabPluto}), viewed in the bulge-fixed frame. The size of the host body, $M_\mathrm{h}$, is varied, and for $M_\mathrm{h}=0.1212$ also the load amplitude is varied. Dotted lines represent models without a fossil bulge. Solid lines show simulations in which a 50 km thick elastic lithosphere has formed only after the hydrostatic shape was reached -- the equilibrium position of the load is shown by solid circles in these cases. The black cross is the sub-host point and the black star is the north pole. a) Positive loads, located initially at 1°E, 89°N. b) Negative loads, initially placed at 1°E, 1°N.}
    \label{fig:Pluto_thereback}
\end{figure*}

We begin the study of Pluto's reorientation with a demonstration of how the $M_\mathrm{h}/M$ ratio affects the paths of positive and negative loads, initially placed near the north pole and the sub-host point, respectively (Fig.~\ref{fig:Pluto_thereback}). Such an exercise is instructive, since most of the previous studies of icy bodies have focused on the case where the host body is much heavier than the orbiting planet or moon ($M_\mathrm{h}/M\rightarrow\infty$).

In this exercise, the loads are placed far away from their equilibrium positions, with the goal to address complete overturns of the body. In each simulation, Pluto is first spinning at angular velocity $\omega_0$ (Table 1) until it reaches an initial equilibrium state. In the first set of simulations (solid lines in Fig.~\ref{fig:Pluto_thereback}), the initial state is assumed to be hydrostatic and the elastic lithosphere to be created only after the equilibrium was established, leading to the formation of a "fossil" bulge \citep[e.g.][]{Matsuyama2014}. In the second set of simulations (dotted lines), the elastic lithosphere is included from the beginning of the simulation so that no fossil bulge is created. Fig.~\ref{fig:Pluto_thereback} shows the load paths predicted for positive (a) and negative (b) loading and different values of $M_\mathrm{h}/M$. The duration of each simulation is 2 Gyr.

\begin{table}
\centering
\begin{tabular}{c c c}
\hline
   \multicolumn{3}{c}{Internal structure of Pluto} \\
   Outer radius & 1188 & km \\
   Ice/water radius & 1038 & km \\
   Water/core radius & 858 & km \\
   Ice density & 950 & kg/m$^3$ \\
   Water density & 1000 & kg/m$^3$ \\
   Core density & 3360 & kg/m$^3$ \\   
   Ice shear modulus & 3.49 & GPa \\
   Surface temperature, $T_\mathrm{surf}$ & 47 & $K$ \\
   Ice/water temperature, $T_\mathrm{bot}$ & 265 & $K$ \\
   Activation energy, $E$ & 59 & kJ \\
   Exponential prefactor, $A$ & $9\cdot 10^{-8}$ & m$^2$K/(Pa s) \\
   Grain size, $d$ & 10 & mm \\
   Cut-off viscosity, $\eta_\mathrm{max}$ & $10^{24}$ & Pa s \\
   Lithosphere thickness, $D_\mathrm{EL}$ & 50 & km \\
   Initial spin rate, $\omega_0$ & $1.14074 {\cdot} 10^{-5}$ & rad/s \\
\hline
   \multicolumn{3}{c}{Parameters of the disc load} \\
   Lateral extent & 25 & ° \\
   Thickness, $h$ & -300 to 300 & m \\
   Density & 1000 & kg/m$^3$ \\
   Growth time, $t_\mathrm{load}$ & 10 & Myr  \\
\hline
\end{tabular}
\caption{Model of Pluto, the parameters of ice shell are taken from \citet{Johnson2016} and \citet{Kihoulou2022b}.}
\label{tabPluto}
\end{table}

A comparison of panels a and b in Fig.~\ref{fig:Pluto_thereback} shows that the load paths for negative loads (Fig.~\ref{fig:Pluto_thereback}b) significantly differ from those for positive loads (Fig.~\ref{fig:Pluto_thereback}a). Even though the initial perturbation in longitude is only 1°, negative loads never move directly towards the north pole, regardless of the value of $M_\mathrm{h}/M$ and of whether the body has a remnant figure or not. 

In case of models without a fossil bulge (dotted lines), the load paths straighten as $M_\mathrm{h}/M$ decreases when the load is positive, while the opposite is true for a negative load.

When a fossil bulge is considered, the load paths obtained for the positive loads are nearly straight. The negative loads first move along the equator and turn to the north only after approaching the center of the trailing hemisphere. This peculiar behaviour is analyzed below. The dependence of the solution on the load size is illustrated for the Charon/Pluto mass ratio ($M_\mathrm{h}/M=0.1212$), showing that the extent of reorientation decreases non-linearly with the load magnitude.

\begin{figure}
    \centering
    \includegraphics[width=0.45\textwidth]{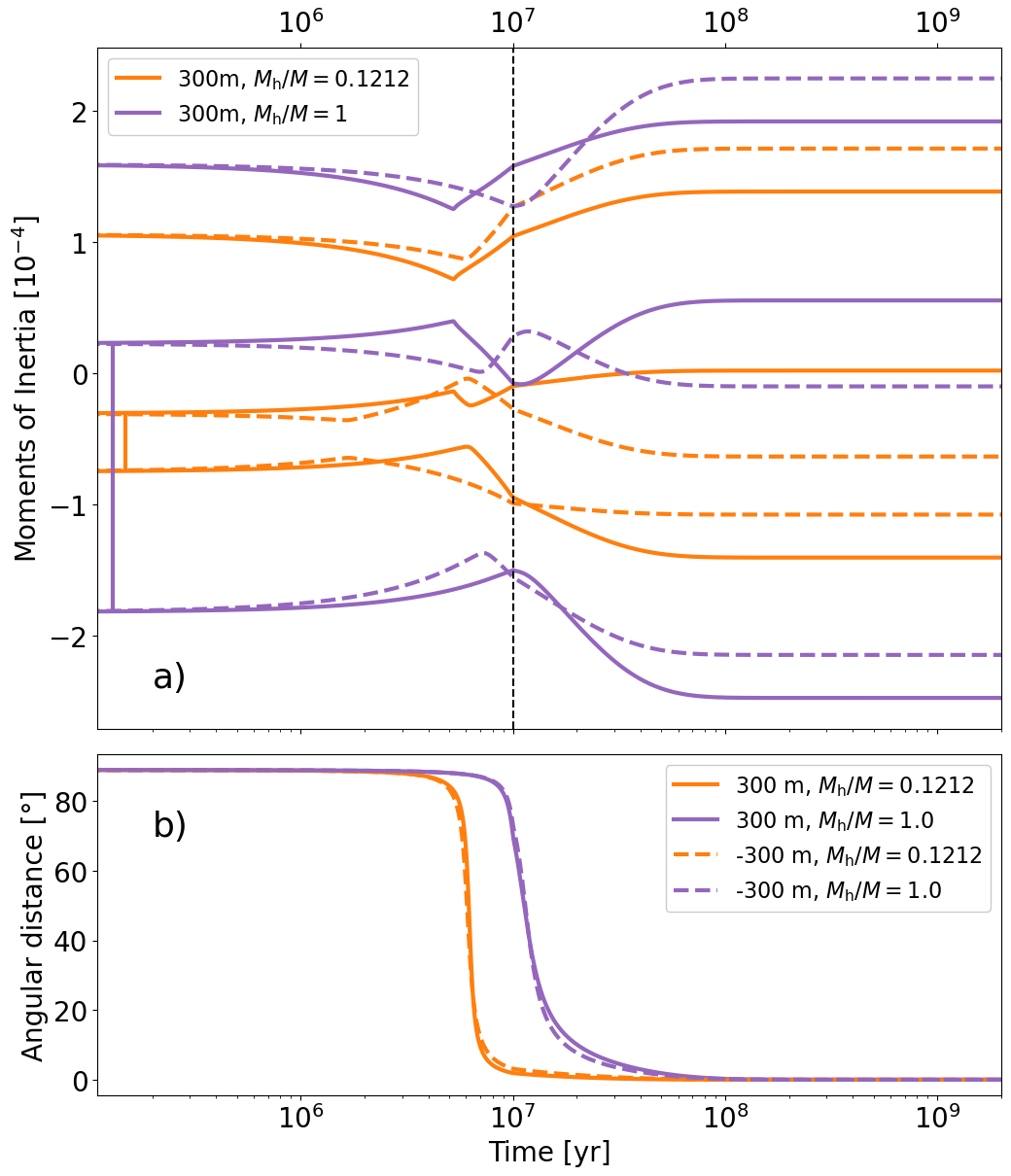}
    \caption{a) Moments of the deviatoric part of the inertia tensor $\bm{I}$ during the TPW simulations plotted by the dotted orange and purple lines in Fig.~\ref{fig:Pluto_thereback} (i.e.~without a fossil bulge). Solid lines are for the positive load of 300 m thickness, dashed lines are for the negative load of the same magnitude. The values are normalized by the diagonal element of the spherical part of the inertia tensor, i.e.~by one third of the trace of $\bm{I}$. Full onset of the load, $t_\mathrm{load}$, is marked by the vertical dashed line. b) The temporal progress of the reorientation. We show the angular distance of the load from its final position for the four cases depicted in panel a.}
    \label{fig:MIA}
\end{figure}

To understand the observed behaviour, one must focus on the way in which the inertia tensor of the load combines with that of the initial figure. While the inertia contribution of the positive load combines with the initial figure such as to overturn the major and the intermediate axes of the total inertia tensor $\bm{I}$, the negative load acts such as to overturn the intermediate and the minor axes. This is expressed by the evolution of the $C-B$ and $B-A$ differences during the growth time of the load (10 Myr, Table 1). When the load is negative, it is $B-A$ that decreases during the first few Myr, while for the positive load the $C-B$ difference narrows. It is thus the value of $B-A$ that stabilizes the longitude of the negative load. When the mass of the host body $M_\mathrm{h}$ is small, the initial figure approaches that of a tidally undeformed body, and the initial values of $A$ and $B$ are close to each other (cf.~the purple and orange vertical line segments in Fig.~\ref{fig:MIA}a). As a result, the negative load is more likely change its longitude as $M_\mathrm{h}/M$ decreases (cf.~the solid brown, purple, orange, and blue circles in Fig.~\ref{fig:Pluto_thereback}b). 

When the load is positive, TPW occurs preferentially along 90° longitude, because reorientation in the plane perpendicular to $\tax$ does not require any readjustment of the tidal bulge. As $M_\mathrm{h}/M$ decreases, this stabilization becomes less important, allowing for the load to reach its equilibrium position in a more straightforward manner. 

The curves in Fig.~\ref{fig:MIA}a are quite complicated, because the evolution of the moments of inertia is affected by several processes: i) the linear rise of the load (full onset is marked by the vertical black dashed line), ii) the progress of TPW itself (Fig.~\ref{fig:MIA}b), and iii) readjustment of the tidal and rotational bulges in response to the TPW.

A more graphic explanation of the behaviour obtained in Fig.~\ref{fig:Pluto_thereback} is shown in Appendix \ref{sec:Janal}, where the principal directions of the equilibrium inertia tensor are plotted with respect to the initial figure of the body. In Fig.~\ref{fig:appendix}, we select a few load amplitudes and analyze how the individual contributions combine to form the total inertia tensor $\bm{I}$.

\begin{figure*}[t]
    \centering
    \includegraphics[width=0.45\textwidth]{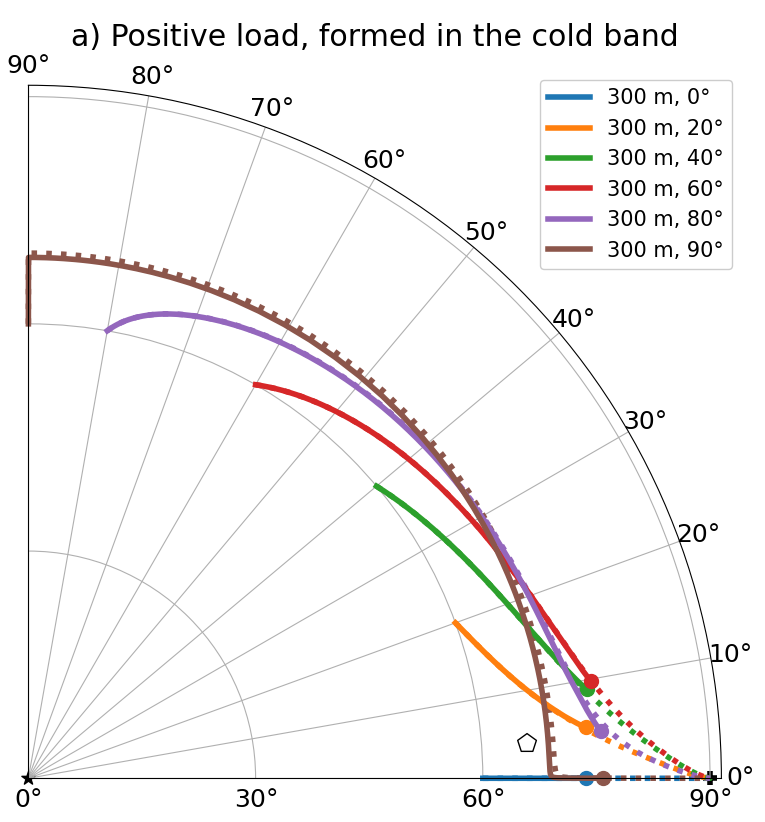}
    \includegraphics[width=0.45\textwidth]{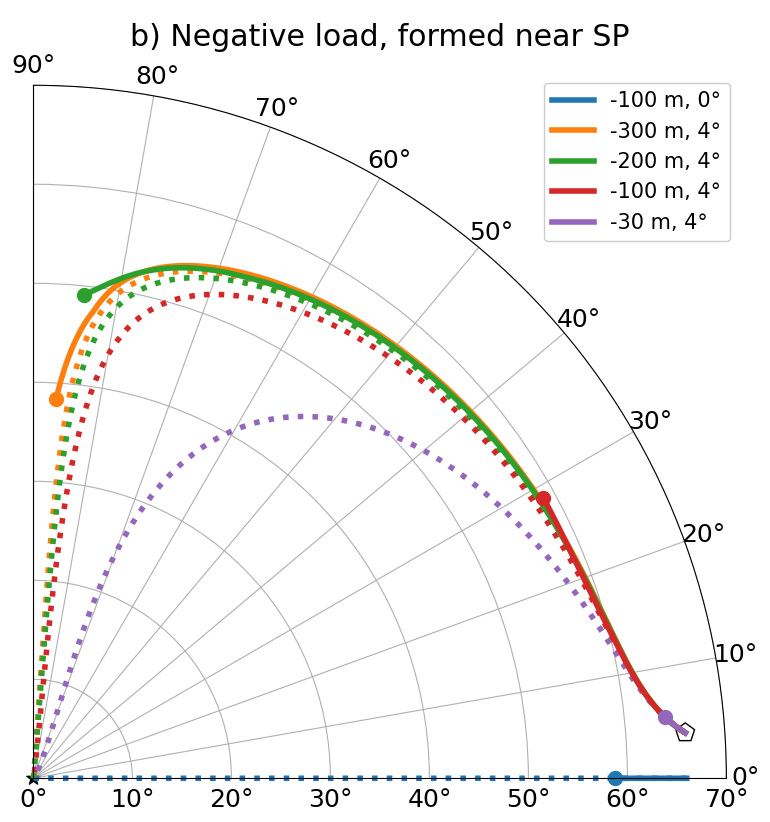}
    \caption{Same as Fig.~\ref{fig:Pluto_thereback}, only here it is the initial position and amplitude of the disc load that is varied. The ratio $M_\mathrm{h}/M$ is 0.1212 for all cases. Black pentagon marks the mirror image of Sputnik Planitia's center, i.e., its position in the first quadrant of polar coordinates. a) Disc thickness is 300 m. Initial longitude of the positive anomaly is taken as 0°, 20°, 40°, 60°, 80°, and 90°E. b) Negative anomaly of a varying magnitude, initially placed at 24°N, 4°E, with the exception of the blue line, for which the disc is placed at 24°N, 0°E.}
    \label{fig:Pluto_Ham}
\end{figure*}

The results in Fig.~\ref{fig:Pluto_thereback}b cast a doubt on the hypothesis of \citet{Kihoulou2022}, who have suggested that Sputnik Planitia first moved to the anti-Charon point, where it resided until the ice shell grew thicker and a fossil bulge was formed, and only then moved towards the north due to the disappearance of the ocean uplift for the thicker ice shell (and thus change of sign of the gravity anomaly). In fact, already a slight perturbation in longitude would trigger a rapid motion of the load in the direction perpendicular to the plane spanned by $\omg$ and $\tax$, making it unlikely for the basin to reach its present geographic location. Note that some perturbation is always required to start a reorientation process from the sub-host (or the anti-host) point, because these points represent the equilibrium states (stable or unstable, depending on the sign of the load). As shown in Fig.~\ref{fig:Pluto_thereback}b, a negative load can move directly from the anti-Charon point to the north, as suggested by \citet{Kihoulou2022}, only if the latitudinal position of the load is disturbed but its longitudinal position remains fixed at exactly 180° (cf.~also the blue line in Fig.~\ref{fig:Pluto_Ham}b).

In Fig.~\ref{fig:Pluto_Ham}, we explore the concept of \citet{Hamilton2016}. Note that we again employ the planes of symmetry of the problem and place Sputnik Planitia at 24°N, 4°E instead of 24°N, 176°E (black pentagon), because we choose to work in the first quadrant of polar coordinates. In Fig.~\ref{fig:Pluto_Ham}b, we investigate what would have happened if the load had first reached its present location and then changed from positive to negative. Inspection of the figure shows that the answer depends on the presence or absence of a fossil bulge and on the size of the load. In the absence of fossil figure (dotted lines), Pluto would undergo a large reorientation, moving Sputnik Planitia far away from its present position within a few hundred Myr. Regardless of its size, the negative load always tends to move to the east, making it unlikely that Sputnik Planitia could remain as close to the 180° meridian as it is at present. A similar trend is found for models with a fossil bulge (solid lines), but in this case, the final position of the basin depends on the disc thickness. The basin is displaced by less than 3° if the load amplitude is 30 m (solid purple line), while the distance between the initial and final positions of the basin is more than 25° for loads with magnitudes exceeding 100 m.

In Fig.~\ref{fig:Pluto_Ham}a, we impose a positive load at the latitude of 30°N (Pluto's coldest region) and calculate its path for different initial longitudes. In the absence of fossil figure (dotted lines), the load always ends up at the anti-Charon point, i.e., 24° away from the present center of Sputnik Planitia. The present position of the basin can be better predicted by models that include the effect of a fossil bulge. For a 300 m thick disc the equilibrium position has a lower latitude than 24°N regardless of the initial longitude, but for a thinner disc the present position of Sputnik Planitia can be reached (cf.~also Fig.~2a in \citet{Keane2016}). Our results suggest that either i) the elastic lithosphere had already been formed when the nitrogen ice began to accumulate, or ii) Pluto has been experiencing a continuous reorientation, while the geographic location of the nitrogen ice does not change due to an ongoing redeposition.

\section{The rate of reorientation}\label{sec:discussion}

The TPW rate depends on the load amplitude, the internal structure and the shape of the fossil figure. As to the internal structure, two parameters are of particular importance: the thickness of the ice shell and the cut-off viscosity $\eta_\mathrm{max}$, representing the effective viscosity of low-temperature ice. To a first approximation, these two parameters determine the characteristic time of viscoelastic readjustment of the tidal and equatorial bulges and thus control the rate of reorientation.

In case of Pluto, neither of these parameters is known with certainty. The ice thickness is estimated to be between 100 and 180 km \citep[][]{Johnson2016, Denton2020, Kihoulou2022b}, while $\eta_\mathrm{max}$ depends on a number of factors such as the overall stress level, composition and deformation history of ice, and is not well constrained by observations.

The simulations presented in section \ref{sec:pluto} (Figs \ref{fig:Pluto_thereback}-\ref{fig:Pluto_Ham}) are computed for $\eta_\mathrm{max}=10^{24}$ Pa s and an ice thickness of 150 km. In this case, the time required to reach the equilibrium orientation varies from a few Myr to 2 Gyr.  When $\eta_\mathrm{max}$ is decreased by two orders of magnitude (Fig.~\ref{fig:robustness}a,b) the rate of reorientation speeds up, with the final orientation being reached within a few tens of Myr. However, the qualitative behaviour does not change. When the ice shell thickness is varied the load paths are even less affected (Fig.~\ref{fig:robustness}c,d).

Note that the changes in load's latitude and longitude can be highly non-linear in time. For instance, the curved segment of the load path depicted in blue in Fig.~\ref{fig:Pluto_thereback}a takes considerably more time than the preceding drop in latitude. When the shell thickness is increased to 300 km, this last stage of reorientation is not completed within the simulation time of 2 Gyr. Due to non-linearities, reaching the equilibrium orientation may require time that is comparable to the age of the solar system in some cases. With a dynamic method, such peculiarities can easily be revealed, but within the fluid-limit framework they would remain unnoticed.

In all the presented simulations, loading is imposed only gradually in time, with the load magnitude rising linearly over the time $t_\mathrm{load}=10$ Myr (Table 1). If $t_\mathrm{load}$ was significantly reduced, the values of $B$ and $A$ would intersect at some point in time, and thus the two principal directions corresponding to the intermediate and minor moments of inertia would instantaneously revert (``flip''). Such an event would turn the rotational dynamics unstable, triggering a 90° wobble that would disturb the body considerably (similarly, the principal directions of inertia could flip due to a crater formed by a sudden impact). The dynamics of wobble on a tidally locked body are outside the scope of the present paper.

\section{Energy Balance}\label{sec:energy}

In this section, we derive an energy conservation law for a tidally locked planet or moon. The derived formula allows a detailed examination of the different types of energy (thermal, elastic, kinetic, rotational, tidal, and gravitational) and we present it in a form that is suitable for testing the accuracy of any TPW solver.

Conservation of energy in a non-inertial, rotating frame can be written as:
\begin{linenomath*}\begin{multline} 
\dt{}\int_{v(t)}\left(\rho\epsilon+\frac{_1}{^2}\rho\bm{v}\cdot\bm{v}\right)\, \mathrm{d}v \\
=\int_{v(t)}\left(-\nabla U-\nabla\Theta-\nabla\Psi-2\bm{\omega}\times\bm{v}- \dt{\omg}\times\bm{r}\right)\cdot\bm{v}\,\rho \dv,
\label{eq_bal}
\end{multline}\end{linenomath*}
where $v(t)$ is the volume of the body at time $t$. The terms in the integral on the left-hand side (LHS) correspond to the specific internal energy \citep[Eqs (20) and (21) in][]{Patocka2018} and the kinetic energy associated with the deformation, respectively, while the integral on the right-hand side (RHS) represents the power of the body forces acting in a rotating system.

It is easy to see that the power delivered by the Coriolis force is zero, as the force is always perpendicular to the velocity vector $\bm{v}$. Since, by definition, the relative angular momentum, $\bm{h}$, in the Tisserand frame vanishes, the total power of the Euler force is also zero:
\begin{linenomath*}\begin{multline}
\int_{v(t)}\left(-\rho\dt{\omg}\times\bm{r}\right)\cdot\bm{v}\,\mathrm{d}v = -\dt{\omg}\cdot\int_{v(t)}\bm{r}\times(\rho\bm{v})\,\mathrm{d}v \\ 
= -\dt{\omg}\cdot\bm{h}\ = \bm{0}. \label{eqEulerPo}
\end{multline}\end{linenomath*}
The power of the centrifugal force is \citep[see Eq.~(A7) in][]{Patocka2021}:
\begin{linenomath*}\begin{multline} 
   \ivo{-\nabla\Psi \cdot\bm{v} \rho\dv} = \ivo{\left[ (\bm{r}{\cdot}\bm{v})\,\omega^2 -(\omg{\cdot}\bm{v})\cdot(\omg{\cdot}\bm{r})\right] \rho\dv} \\
   = \frac{1}{2}\omg\cdot \dt{\bm{I}}\cdot\omg. \label{eqCentrPo}
\end{multline}\end{linenomath*}
Using Eq.~\eqref{eqCentrPo} and taking into account that the power $-\nabla U \cdot \bm{v}$ that is associated with the deformation of the body in its own gravitational potential, $U$, is provided the gravitational energy $\int \rho U/2 \dv$ \citep[Eq.~(A21) in][]{Patocka2018}, Eq.~\eqref{eq_bal} can be rewritten as follows:
\begin{linenomath*}\begin{multline}
\dt{}\int_{v(t)}\left(\epsilon+\frac{_1}{^2}\bm{v}\cdot\bm{v} + \frac{1}{2} U\right)\,\rho\, \mathrm{d}v \\
=\frac{1}{2}\omg\cdot \dt{\bm{I}}\cdot\omg - \ivo \nabla \Theta\cdot\bm{v} \rho\dv . \label{eq_balrot}
\end{multline}\end{linenomath*}
Adding the rate of the rotational energy, $E_\mathrm{rot}=(\omg{\cdot} \bm{I}{\cdot}\omg)/2$, to both sides, we get:
\begin{linenomath*}\begin{align}
\dt{}\int_{v(t)}\left(\epsilon+\frac{1}{2}\bm{v}\cdot\bm{v} +\frac{1}{2} U - \Psi\right)\,\rho\,\mathrm{d}v \nonumber\\ = \frac{1}{2}\omg\cdot \dt{\bm{I}}\cdot\omg + \dt{} \left( \frac{1}{2}\omg\cdot \bm{I}\cdot\omg \right) - \ivo \nabla \Theta\cdot\bm{v} \rho\dv. \label{eq_balrot2}
\end{align}\end{linenomath*}
Note that the rotational energy is obtained by integrating $-\Psi$, while in the case of the gravitational energy the integrand is $+U/2$. Combining the first two terms on the RHS and using the symmetry of $\bm{I}$, we get:
\begin{linenomath*}\begin{multline}
\dt{}\left(E_\mathrm{int} + E_\mathrm{kin} + E_\mathrm{grav} + E_\mathrm{rot}\right) \\ = \dt{( \bm{I} \cdot \omg )} \cdot\omg - \ivo \nabla \Theta\cdot\bm{v} \rho\dv. \label{eq_balrot3a}
\end{multline}\end{linenomath*}
The first term on the RHS can be expressed in terms of the torque $\bm{M}$, because the projection of the LE onto $\omg$ reads:
\begin{linenomath*}\begin{equation}\label{eq_LEdotw}
\bm{M} \cdot \omg = \left( \dt{( \bm{I} \cdot \omg )} + \omg\times(\bm{I}\cdot\omg)\right) \cdot\omg = \dt{( \bm{I} \cdot \omg )} \cdot\omg.
\end{equation}\end{linenomath*}
Eq.~\eqref{eq_balrot2} then takes the form
\begin{linenomath*}\begin{multline}
\dt{}\left(E_\mathrm{int} + E_\mathrm{kin} + E_\mathrm{grav} + E_\mathrm{rot}\right) \\ = \bm{M} \cdot\omg - \ivo \nabla \Theta\cdot\bm{v} \rho\dv. \label{eq_balrot3b}
\end{multline}\end{linenomath*}
Since Eq.~\eqref{eq_balrot3b} is valid if and only if the LE is satisfied, it can be used to test the accuracy of the TPW solution: For a tide-free body (i.e., $\bm{M}=0$ and $\Theta=0$), the sum of $E_\mathrm{int} + E_\mathrm{kin} + E_\mathrm{grav} + E_\mathrm{rot}$ must be conserved. As shown by \citet{Patocka2021}, this is satisfied only if the changes in $\Psi$ are treated in a self-consistent manner and to a high degree of precision in the course of deformation.

If the body is subject to tidal forcing, the RHS in Eq.~\eqref{eq_balrot3b} is non-zero, implying that the sum of the energies on the LHS is no more conserved. Analogously to Eq.~\eqref{eqCentrPo}, the second term on the RHS can be expressed as
\begin{linenomath*}\begin{equation}\label{eqTidPo}
    \ivo \nabla \Theta\cdot\bm{v}\, \rho \dv  = \frac{1}{2}\tax\cdot \dt{\bm{I}}\cdot\tax + \dt{}\ivo \frac{1}{3} r^2 \taxa^2 \rho \, \dv.
\end{equation}\end{linenomath*}
For convenience, we define the ``tidal'' energy
\begin{linenomath*}\begin{equation}\label{eqTidEn}
    E_\mathrm{tid} = \ivo \Theta \rho\,\dv = \frac{1}{2} \tax\cdot\bm{I}\cdot\tax - \ivo \frac{1}{3} r^2 \taxa^2 \,\rho\,\dv.
\end{equation}\end{linenomath*}
The tidal energy is the potential energy possessed by the body due to the gravity of the host body. Therefore, $\Theta$ is taken with a positive sign in Eq.~\eqref{eqTidEn}, but the usual factor of 1/2 is omitted. Recalling that $\bm{M}=\tax\times(\bm{I}\cdot\tax)$ and using the identity $\bm{a}\cdot(\bm{b}\times\bm{c}) = \bm{c}\cdot(\bm{a}\times\bm{b})$, the energy balance, Eq.~\eqref{eq_balrot3b}, can be rearranged as follows:
\begin{linenomath*}\begin{multline}
\dt{}\left( E_\mathrm{int} + E_\mathrm{kin} + E_\mathrm{grav} + E_\mathrm{rot} + E_\mathrm{tid} \right) \\ =(\bm{I}\cdot\tax)\cdot \left(\omg\times\tax + \dt{\tax} \right) . \label{eq_balrot3}
\end{multline}\end{linenomath*}
The RHS of Eq.~\eqref{eq_balrot3} describes the energy that is exchanged with the host body throughout the reorientation process.

Note that in deriving Eq.~\eqref{eq_balrot3}, no specific assumptions have been made about the form of the LE. Under the $\tomap$ approximation, Eq.~\eqref{eq_balrot3} reduces to:
\begin{linenomath*}\begin{equation}\label{eq_balwMIA}
\dt{}\left( E_\mathrm{int} + E_\mathrm{kin} + E_\mathrm{grav} + E_\mathrm{rot} + E_\mathrm{tid} \right)= A\taxa\dt{\taxa},
\end{equation}\end{linenomath*}
where $A$ is the minimum moment of inertia and the RHS can be evaluated from the solution of Eqs~\eqref{eqtomap_1}--\eqref{eqtomap_3} by using the relationship between $\omega$ and $\taxa$, Eq.~\eqref{eqTidlock}.

In reality, reorientation may lead to small deviations from the $\tomap$ assumption, but these deviations are likely to be small because they are effectively dampened by dissipative processes.

\begin{figure}
    \centering
    \includegraphics[width=0.5\textwidth]{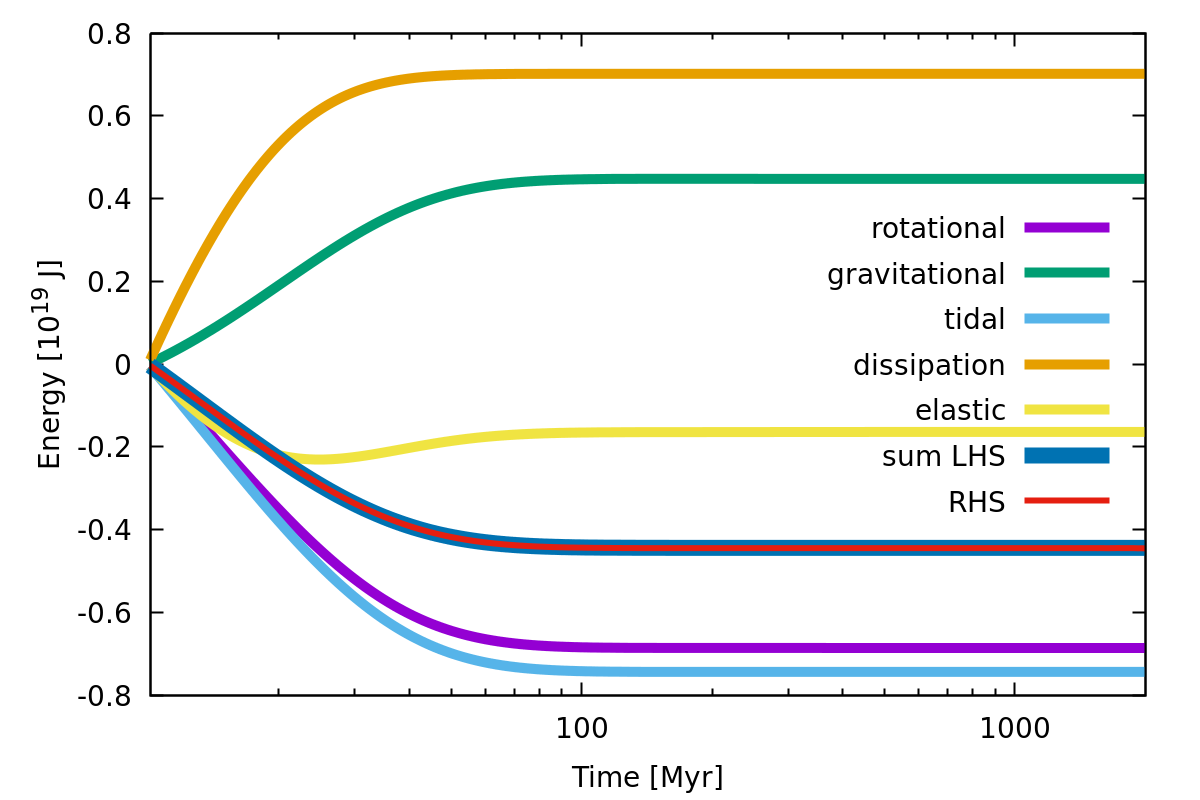}
    \caption{Contributions to the total energy and their evolution for model with $M/M_\mathrm{h}=0.1212, h=300$ m and no fossil bulge. The terms in Eq.~\eqref{eq_balwMIA} are integrated over time and shown in different colors. The initial time of the plot $t=10$ Myr corresponds to the full onset of loading. $E_\mathrm{kin}$ is negligible and thus not shown for figure clarity.}
    \label{fig:energy}
\end{figure}

The use of the energy balance, Eq.~\eqref{eq_balwMIA}, is illustrated in Fig.~\ref{fig:energy}.
Inspection of the figure shows that most types of energy included in Eq.~\eqref{eq_balwMIA} are of the same order of magnitude and the LHS (dark blue line) is equal to the RHS (thin red line), confirming that the numerical solution is correctly implemented. The relative changes of the spin rate in this simulation are of the order $10^{-5}$, which corresponds to a change in the Pluto-Charon distance of $\sim 100$ m. Note that the rotational and tidal energies are minimized during the reorientation, while the gravitational energy of the disc load in the gravitational potential of the studied body is increased, because the disc is lifted against the surface gravity by the readjustment of the rotational and tidal bulges.

\section{Summary}\label{sec:Sum}

We have developed a new method to investigate the dynamics of reorientation of a tidally locked body. Our method gives the same results as that of \citet{Hu2017a}, but is much simpler, requiring only the evaluation of the eigenvalues and eigenvectors of the inertia tensor. The theoretical framework of the method is completed by an analysis of the energy balance, which can be used to test the numerical solution. A new release of the code LIOUSHELL that was used to perform the simulations is freely available on GitHub \citep{Patocka_LIOUSHELL20}.

Unlike most studies that only investigate the final (equilibrium) orientation of the body \citep[recently, e.g.,][]{Schenk2020,Matsuyama2021,Johnson2021}, the $\tomap$ method presented here can be used to predict the wander of the north pole and the sub-host point, i.e., the evolution of the rotation and tidal vectors in the body-fixed frame, or the path of the surface load in the geographic frame. Since the rate of reorientation depends on the internal structure and material parameters of the body, the modeling of reorientation dynamics can be used to explore the properties of the body and to test different scenarios of its evolution. 

We investigate the path of a positive load placed near one of the rotation poles and the path of a negative load placed near the tidal axis, varying the mass ratio $M_\mathrm{h}/M$ where $M$ is the mass of the body under consideration and $M_\mathrm{h}$ is the mass of the body whose gravitational pull generates the tidal force. When $M_\mathrm{h}/M$ decreases, the load path straightens for the positive load and becomes more curved for the negative load. The negative load tends to rapidly change its longitude when $M_\mathrm{h}/M$ is small, because then the equatorial moments of inertia are close in value.
  
The asymmetric response of a tidally locked body to positive and negative loads has important consequences for the reorientation of Pluto ($M/M_\mathrm{h}=0.12$) and the present-day location of the Sputnik Planitia basin. In particular, we show that it is unlikely that the load associated with Sputnik Planitia changed sign from positive to negative after reaching the anti-Charon point, as proposed by \citet{Kihoulou2022}. A more plausible scenario is that the basin represents a positive load and its center stays away from the equator due to a fossil bulge that existed already at the time of Sputnik Planitia's formation \citep{Keane2016}. If it is the uplift of a dense interior ocean that compensates the negative topography of the basin \citep{Nimmo2016}, then the ice shell of Pluto must be less than 200 km thick and insulated by a layer of clathrates at its bottom \citep{Kamata2019,Kihoulou2022b}. Our results do not support the hypothesis that Sputnik Planitia was formed early in Pluto's history \citep{Hamilton2016}, because in the absence of fossil figure the basin would move directly to the anti-Charon point.

\section*{Acknowledgements}

V.P.~and M.K.~acknowledge support by the Czech Science Foundation through project nr.~22-20388S.

\printcredits
\appendix
\renewcommand\thefigure{\thesection.\arabic{figure}}

\section{Analysis of the Inertia Tensor Constituents}\label{sec:Janal}
\setcounter{figure}{0}

In the initial state of our simulations with a fossil bulge, the inertia tensor of the model Pluto can be written as $\bm{I} = \bm{I}_\mathrm{sph} + \bm{I}_\mathrm{hyd}$, where $\bm{I}_\mathrm{sph}$ is the inertia tensor of the undeformed model (a sphere with a prescribed density profile), and $\bm{I}_\mathrm{hyd}$ is the hydrostatic deformation due to the centrifugal and tidal potentials. Fossil bulge, $\bm{I}_\mathrm{foss}$, is then the part of the hydrostatic figure that is frozen into the cold, elastic lithosphere, and does not readjust when the direction of $\omg$ or $\tax$ changes \citep[it is computed as described in section 6 of][]{Patocka2021}. The inertia contribution of the load that is gradually added to the body within the time $t_\mathrm{load}$ is denoted as $\bm{I}_\mathrm{disc}$.

The inertia tensors $\bm{I}_\mathrm{foss}$ and $\bm{I}_\mathrm{disc}$ represent deformations of the body, not additional masses. As such, they are traceless, that is, the sum of their eigenvalues is zero. For any such tensor, two situations can occur: i) The eigenvalue with the largest magnitude is a positive number. This means that the dominant deformation is a flattening along the respective eigenvector (such as the flattening caused by the centrifugal potential along the vector $\omg$). In this case, we plot the tensor as an ellipse, whose long and short axes correspond respectively to the minor and intermediate eigenvalues of the tensor (this ellipse becomes a circle that represents the equatorial ring in case of a rotationally flattened body). ii) The eigenvalue with the largest magnitude is negative. This means that the dominant deformation is a stretching along the respective eigenvector (such as the pulling along $\tax$ that is caused by the tidal force). In this case, we plot the tensor as a pair of outward arrows (the upper arrow points towards the center of the disc when $\bm{I}_\mathrm{disc}$ of a positive load is plotted). If the two remaining eigenvalues are not the same, we complement the outward arrows with a pair of inward arrows that show the major axis of the tensor.

The equilibrium (final) orientation of the body is obtained by diagonalizing $\bm{I}_\mathrm{foss} + \bm{I}_\mathrm{disc}$, with the equilibrium direction of $\omg$ and $\tax$ lying respectively along the major and minor axes of the combined tensor (note that $\bm{I}_\mathrm{foss} + \bm{I}_\mathrm{disc}$ is traceless, and thus the major axis is the one whose eigenvalue is the largest in value, but not necessarily the largest in the absolute value). In our graphic representation, this means that the equilibrium direction of $\omg$ is perpendicular to the ellipse (when $\bm{I}_\mathrm{foss} + \bm{I}_\mathrm{disc}$ is plotted as an ellipse), or lies along the inward pointing arrows when $\bm{I}_\mathrm{foss} + \bm{I}_\mathrm{disc}$ corresponds to a pulled rather than to a flattened sphere. Similarly, $\tax$ lies either along the long axis of the ellipse, or along the outward pointing arrows.

\begin{figure*}[t]
    \centering
    \includegraphics[trim=0 30 0 0, clip,width=0.45\textwidth]{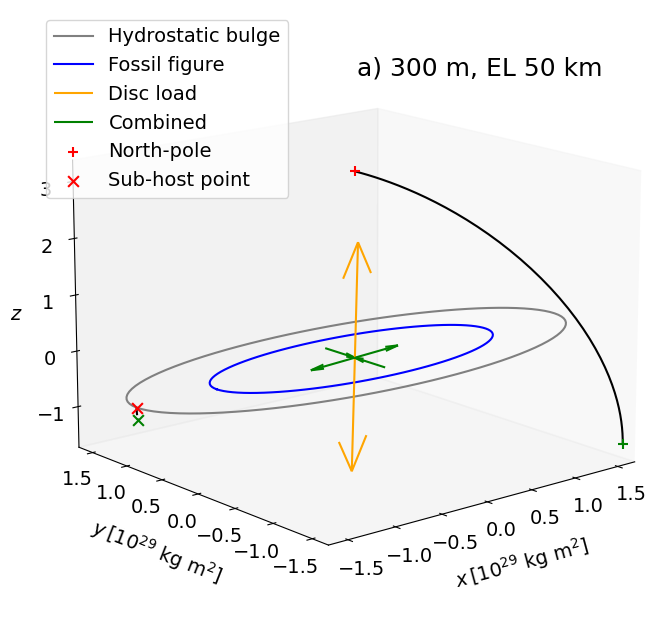}
    \includegraphics[trim=0 30 0 0, clip,width=0.45\textwidth]{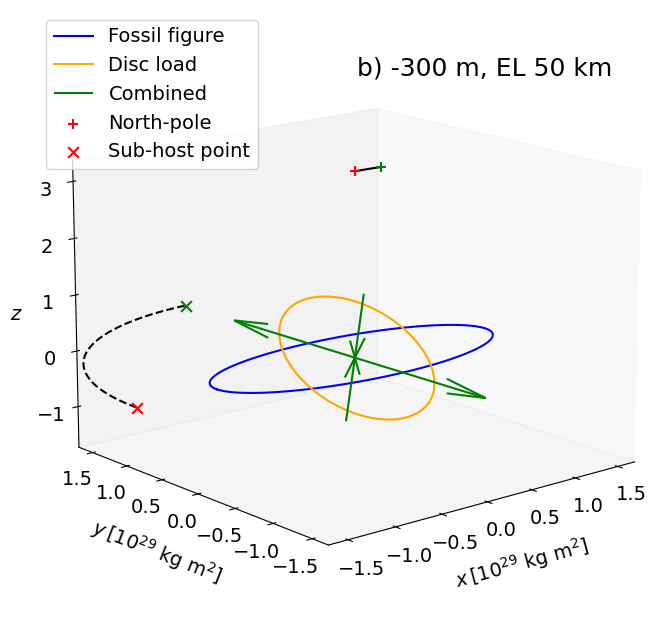}
    \includegraphics[trim=0 30 0 0, clip,width=0.45\textwidth]{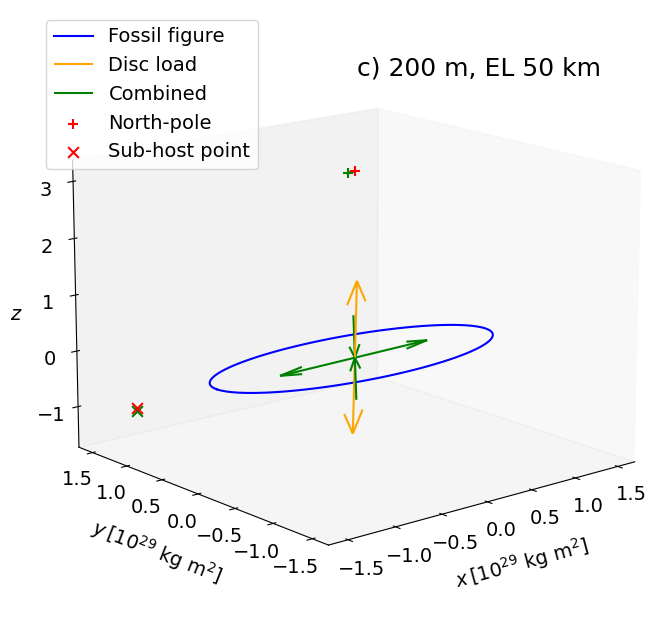}
    \includegraphics[trim=0 30 0 0, clip,width=0.45\textwidth]{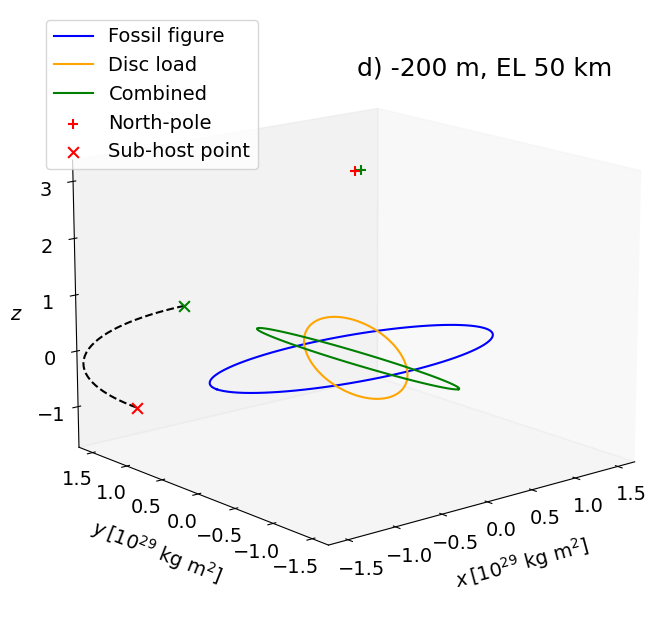}
    \includegraphics[trim=0 30 0 0, clip,width=0.45\textwidth]{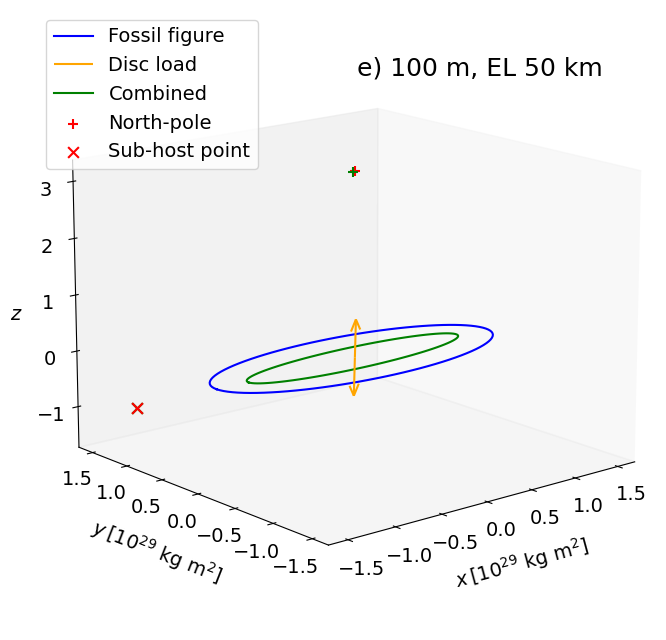}
    \includegraphics[trim=0 30 0 0, clip,width=0.45\textwidth]{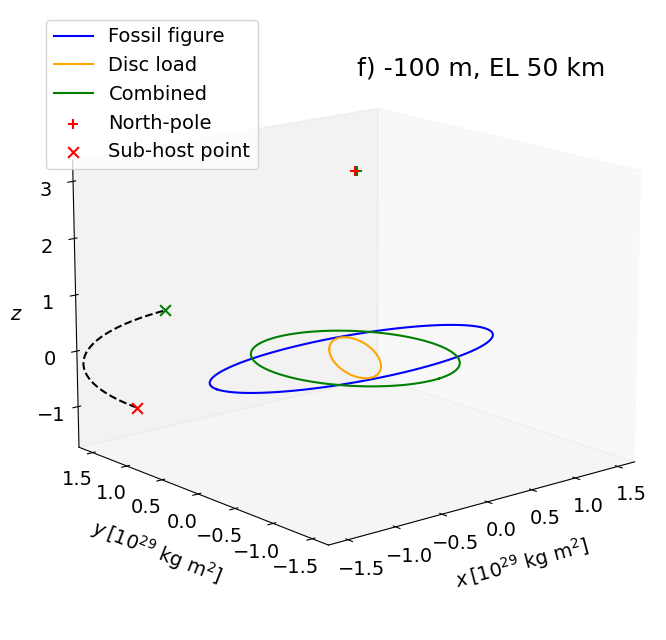}
    \caption{Contributions to the planet's inertia tensor at the time of loading, plotted in the body-fixed frame. North pole's initial position is marked by the red plus sign, its equilibrium position is marked by the green plus sign (the two are connected with a black solid line). The equilibrium north pole lies along the major direction of the combined inertia tensor (green), i.e.~that of the imposed disc (yellow) summed with that of the fossil bulge (blue). The initial and the equilibrium positions of the sub-host point (red and green crosses) are connected with a black dashed lines. See text for an explanation of the way the tensors are represented. a),c),e) Model Pluto with a remnant figure corresponding to a 50 km thick lithosphere, loaded by the disc with a thickness of respectively 300, 200, and 100 m. b),d),f) The same, only here for the negative amplitudes. }
    \label{fig:appendix}
\end{figure*}

In Fig.~\ref{fig:appendix}, we show $\bm{I}_\mathrm{foss}$ (in blue colour), $\bm{I}_\mathrm{disc}$ (yellow), and the combined tensor $\bm{I}_\mathrm{foss} + \bm{I}_\mathrm{disc}$ (green) for the six simulations that are plotted by the solid orange, green, and red lines in Fig.~\ref{fig:Pluto_thereback}a,b (i.e., $M/M_\mathrm{h}=0.1212$, with a fossil bulge, the load amplitude varies). Before the onset of loading, the body is in the hydrostatic shape (see $\bm{I}_\mathrm{hyd}$ in Fig.~\ref{fig:appendix}a, note that $M/M_\mathrm{h}=0.1212$ and thus $\bm{I}_\mathrm{hyd}$ is depicted as an ellipse, because the rotational flattening is stronger than the tidal pull). The initial north pole is connected with the equilibrium north pole by a black solid line in the body-fixed frame, the initial and the equilibrium sub-host points are connected with a black dashed line. The principal directions of the combined inertia tensor depend on $\bm{I}_\mathrm{foss}$ and $\bm{I}_\mathrm{disc}$ in a non-trivial, highly non-linear way, as illustrated by the various possibilities in the different panels of Fig.~\ref{fig:appendix}.

In particular, the negative load acts such as to switch the long and the short axes of $\bm{I}_\mathrm{foss}$, and this can be successfully done even when $|\bm{I}_\mathrm{disc}|$ is significantly smaller than $|\bm{I}_\mathrm{foss}|$ (panel f). As a result, the load is very likely to change its equilibrium longitude (see the dashed lines in panels b,d, and f). The positive load, on the other hand, does not act to switch the long and the short axes of $\bm{I}_\mathrm{foss}$ (the load symmetry lies in the direction that is nearly perpendicular to the blue ellipse), and thus the equilibrium displacement of $\tax$ is always small. When sufficiently large, however, the positive load can stretch the body such as to overturn the major axis of the fossil figure, causing a large displacement of $\omg$ in effect (the solid black line in panel a).

\section{Sensitivity to $\eta_\mathrm{max}$ and ice shell thickness} \label{sec:robustness}
\setcounter{figure}{0}

\begin{figure*}[t]
    \centering
    \includegraphics[width=0.45\textwidth]{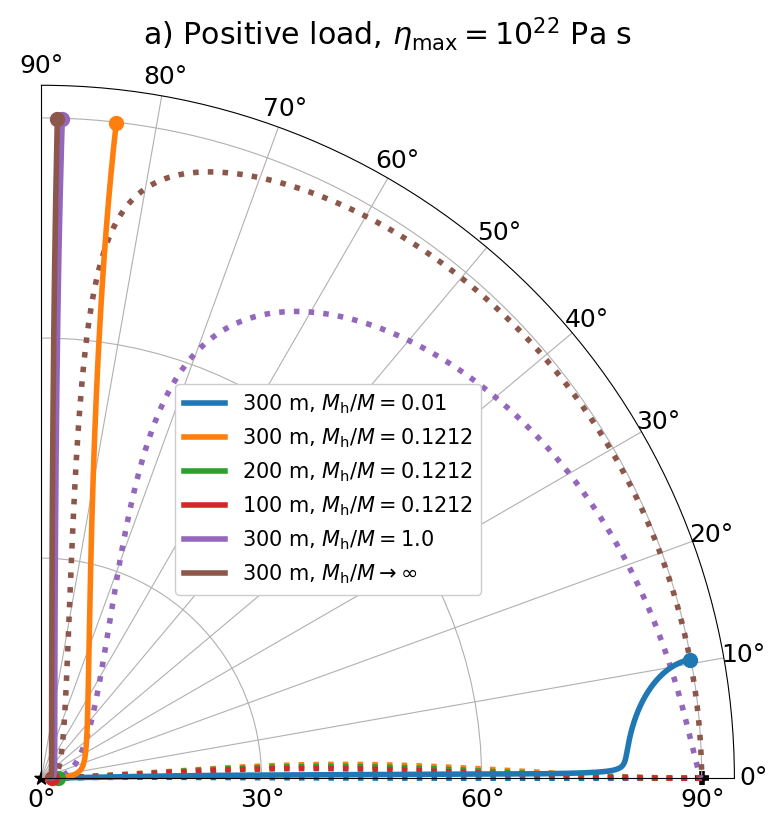}
    \includegraphics[width=0.45\textwidth]{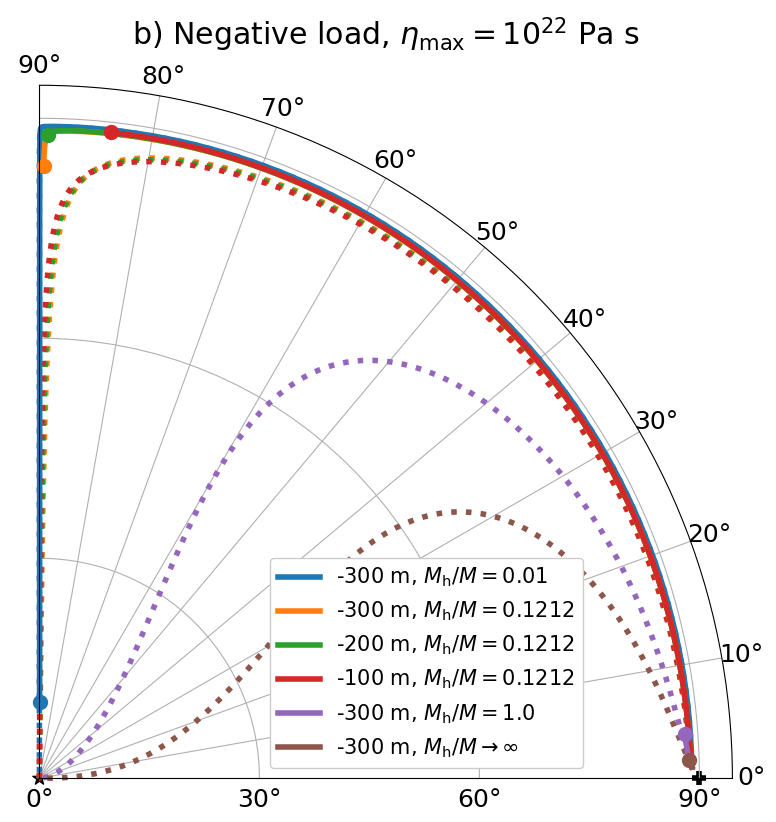}
    \includegraphics[width=0.45\textwidth]{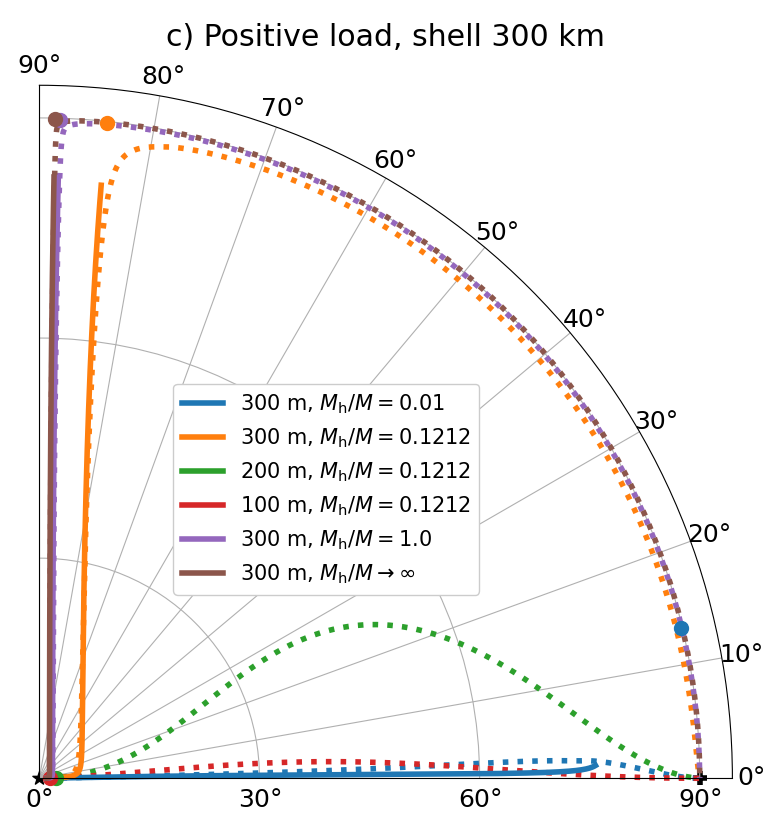}
    \includegraphics[width=0.45\textwidth]{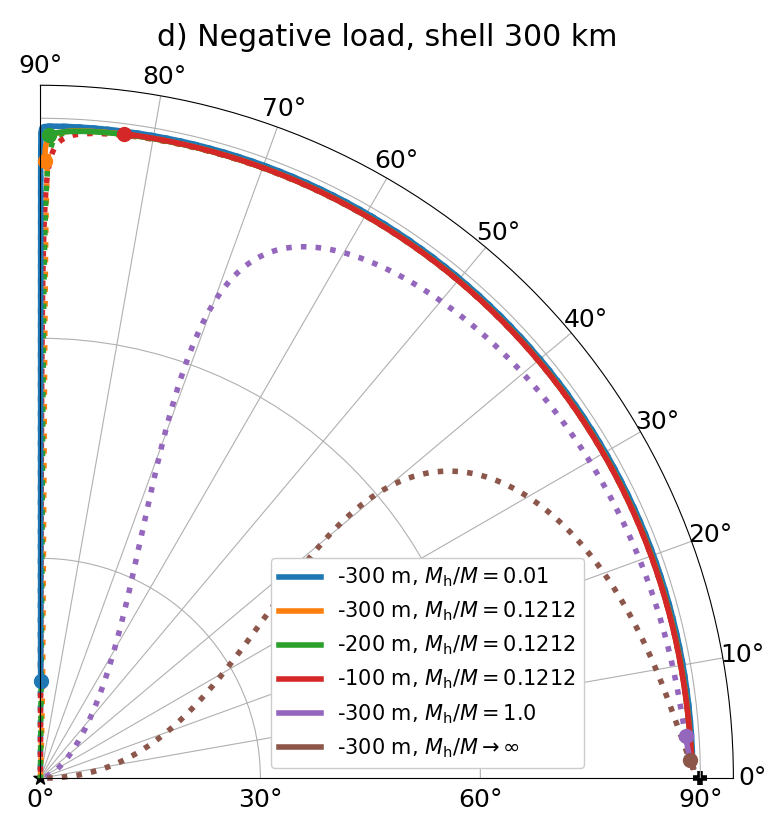}
    \caption{Same as Fig.~\ref{fig:Pluto_thereback}, only here $\eta_\mathrm{max} = 10^{22}$ instead of $10^{24}$ Pa s is considered in the top panel, and a 300 instead of 150 km thick ice shell is assumed in the bottom panel. For a better comparison, the position of the figure legend is kept fixed in all cases. In panel c), the equilibrium positions are not reached within the simulation time of 2 Gyr.}
    \label{fig:robustness}
\end{figure*}

\bibliographystyle{cas-model2-names}

\bibliography{refs}

\end{document}